\documentclass[journal]{IEEEtran}
\usepackage{cite}
\usepackage{bm}
\usepackage{url}
\usepackage{amsmath}
\usepackage{amssymb}
\usepackage[ruled,linesnumbered]{algorithm2e}
\usepackage{diagbox}
\usepackage{tikz}
\usepackage{pgfplots}
\usepackage{graphicx}
\usepackage{subcaption}
\usepackage{multirow}
\usepackage{booktabs}
\usepackage{caption}

\pgfplotsset{compat=1.18}
\begin{document}
	
	\title{Approximate Message Passing-Enhanced Graph Neural Network for OTFS Data Detection}
	\author{Wenhao Zhuang,
		Yuyi Mao, \emph{Member, IEEE}, 
		Hengtao He, \emph{Member, IEEE}, Lei Xie, \emph{Member, IEEE},\\
		Shenghui Song, \emph{Senior Member, IEEE}, Yao Ge, \emph{Member, IEEE}, and Zhi Ding, \emph{Fellow, IEEE}
		\thanks{W. Zhuang and Y. Mao are with the Department
			of Electrical and Electronic Engineering, Hong Kong Polytechnic University, Hong Kong (e-mails: \{wen-hao.zhuang, yuyi-eie.mao\}@polyu.edu.hk). H. He, L. Xie, and S. Song are with the Department of Electronic and Computer Engineering, Hong Kong University of Science and Technology, Hong Kong (e-mails: \{eehthe, eelxie, eeshsong\}@ust.hk). Y. Ge is with the Continental-NTU Corporate Lab, Nanyang Technological University, Singapore (e-mail: yao.ge@ntu.edu.sg). Z. Ding is with the Department of Electrical and Computer Engineering, University of California at Davis, Davis, CA, USA (e-mail: zding@ucdavis.edu). \emph{(Corresponding author: Yuyi Mao)}}
	}
	\maketitle
	
	\begin{abstract}
		Orthogonal time frequency space (OTFS) modulation has emerged as a promising solution to support high-mobility wireless communications, for which, cost-effective data detectors are critical. Although graph neural network (GNN)-based data detectors can achieve decent detection accuracy at reasonable computational cost, they fail to best harness prior information of transmitted data. To further minimize the data detection error of OTFS systems, this letter develops an AMP-GNN-based detector, leveraging the approximate message passing (AMP) algorithm to iteratively improve the symbol estimates of a GNN. Given the inter-Doppler interference (IDI) symbols incur substantial computational overhead to the constructed GNN, learning-based IDI approximation is implemented to sustain low detection complexity. Simulation results demonstrate a remarkable bit error rate (BER) performance achieved by the proposed AMP-GNN-based detector compared to existing baselines. Meanwhile, the proposed IDI approximation scheme avoids a large amount of computations with negligible BER degradation.
		
	\end{abstract}
	
	
\section{Introduction}
	The sixth-generation (6G) wireless networks are envisioned to support high-mobility communications at a speed of up to 1000 km\slash h \cite{IMT2030}, where the traditional orthogonal frequency  division multiplexing (OFDM) modulation may fail to work due to the severe Doppler effects. To mitigate this challenge, orthogonal time frequency space (OTFS) modulation has been proposed, which modulates data symbols in the delay-Doppler (DD) domain. Hence, the time-varying channels are effectively transformed to their quasi-static equivalence in the DD domain, and  full time-frequency (TF) diversity is exploited \cite{otfs_fundemantals}.
	
	To unleash the potential of OTFS modulation, many data detectors have been developed. Specifically, by exploiting the intrinsic DD-domain channel sparsity, a message passing (MP) algorithm was developed in~\cite{Raviteja2018}, where a sparse factor graph of transmitted symbols was constructed to iteratively compute their posterior probability distributions. However, this algorithm may not converge when channel sparsity decreases. Besides, the approximate message passing (AMP) algorithm was adopted in~\cite{amp_otfs} to design a low-complexity data detector for OTFS systems. Nevertheless, it performs sub-optimally as the DD-domain channel matrices are not independent and identically distributed (i.i.d.) sub-Gaussian. Although the unitary AMP algorithm \cite{Zhengdao_UAMP} fits arbitrary channel matrices, it is hard to be implemented due to the need to perform channel matrix singular value decomposition.
	
	Recently, model-driven deep learning (DL) has drawn growing interest, which synergizes DL techniques with traditional signal processing algorithms to boost the data detection accuracy at affordable computational overhead~\cite{Hengtao_model_driven,Xufan_gcn}. 
	In particular, the graph neural network (GNN)-based detector for OTFS modulation~\cite{Xufan_gcn}, which builds upon a pair-wise Markov random field (MRF) to leverage structural system information, outperforms the MP-based and AMP-based detectors by notable margins. Nonetheless, since the GNN parameters are learned from training data, prior information of transmitted symbols cannot be best utilized, leaving a gap for improvement~\cite{ep_gnn,ipe_gnn,Hengtao_GNN-MIMO}. Also, without dedicated optimization, the inter-Doppler interference (IDI) symbols in OTFS systems result in a complex pair-wise MRF and bring heavy computation to the GNN-based detector. Noticed that AMP~\cite{Bayati_AMP} is powerful in removing inter-symbol interference by progressively refining prior information of transmitted symbols, an AMP-GNN network was developed in~\cite{Hengtao_GNN-MIMO} for massive multiple-input multiple-output detection. Although the AMP-GNN network enjoys the benefits of both AMP and GNN, when being myopically applied for OTFS data detection, the computational complexity remains an obstacle.
	
	In this letter, we propose a novel AMP-GNN-based detector for OTFS modulation, where an AMP and a GNN module collaborate to enhance data detection accuracy by exchanging intermediate estimation results. To achieve cost-effective data detection, a learning-based IDI approximation scheme is developed, which simplifies the pair-wise MRF of OTFS systems that largely determines the complexity of the GNN module. Simulation results show that the proposed detector outperforms the baselines by 41.4$\sim$80.7\% in bit error rate (BER) 
	at $20$ dB signal-to-noise ratio (SNR) with 16-quadrature amplitude modulation (QAM) and approaches a performance upper bound without IDI approximation at 64.2$\sim$67.0\% reduced processing cost.

	\textbf{Notations:} We use boldface lower-/upper-case letters to denote vectors/matrices, and denote the $i$-th column of matrix $\mathbf{R}$ as $\mathbf{r}_{i}$. The ``$\text{vec}$'' operator vectorizes an $B\times A$ array $\{r\left[k, l\right]\}$ as $\mathbf{r}\triangleq \text{vec}\left(\{r\left[k,l\right]\}\right)$, where $r[k,l]$ corresponds to the $(kA+l)$-th dimension of $\mathbf{r}$, i.e., $r_{kA+l}$. Besides, $[\cdot]_C $, $\lfloor \cdot \rfloor$, and $\delta\left(\cdot\right)$ denote the modulo-$C$ operation, floor function, and Dirac delta function, respectively. Moreover, the real and complex Gaussian distribution with mean $\bm{\mu}$ and covariance matrix $\bm{\Sigma}$ are denoted as $\mathcal{N}\left(\bm{\mu},\bm{\Sigma}\right)$ and $\mathcal{CN}\left(\bm{\mu},\bm{\Sigma}\right)$, respectively.
	
	\section{System Model}
	We consider a single-user OTFS system with $M$ subcarriers. In each frame, $MN$ symbols from a $|\mathcal{Q}|$-QAM constellation are first multiplexed in the DD domain over $N$ time slots at the transmitter, denoted as $\{\overline{x}\left[k, l\right]\} \in \mathcal{Q}^{MN}$. Then, the multiplexed symbols are spread over the TF domain by the inverse symplectic finite Fourier transform (ISFFT). Finally, symbols in the TF domain are converted to a continuous waveform through the Heisenberg transform, where a rectangular pulse-shaping filter is adopted. At the receiver, the reverse operations, including the Wigner transform and the SFFT, are performed to obtain the received signal $\{\overline{y}\left[k, l\right]\}\!\in\!\mathbb{C}^{MN}$ in the DD domain, which acts as input of data detector. Next, we detail the channel and signal models.
	\subsection{Channel Model}
	A time-varying wireless channel with $P$ propagation paths is considered, where the channel gain, delay, and Doppler shift associated with the $p$-th path are denoted as $\overline{h}_p$, $\tau_p$, and $\nu_p$, respectively. Denote the subcarrier-spacing as $\Delta f$ and symbol duration as $\Delta T$. The channel impulse response in the DD domain can be expressed as follows \cite{otfs_fundemantals}:
	\begin{align}
		h(\tau, \nu) = \sum_{p=1}^{P} \overline{h}_p \delta(\tau-\tau_p) \delta(\nu-\nu_p), \label{eq:channel}
	\end{align}
	where $\tau_p = \frac{l_p}{M\Delta f}$, $\nu_p = \frac{k_p+\kappa_p}{N\Delta T}$ with $l_p$ and $k_p$ indicating the integer delay and Doppler shift index, respectively, and $\kappa_p \in  [-1\slash2, 1\slash2]$ denoting the fractional Doppler shift.
	\subsection{Signal Model}
	Given the channel model in \eqref{eq:channel}, the received signal in the DD domain is expressed as follows \cite{Raviteja2018}:
	\begin{align}
		\overline{y}[k,l] =& \frac{1}{N} \sum_{p=1}^{P} \overline{h}_p e^{j2\pi \frac{(l-l_{p})(k_p+\kappa_p)}{MN}} \sum_{q=\lfloor-\frac{N}{2}\rfloor+1}^{\lfloor\frac{N}{2}\rfloor} \alpha_p(k,l,q) \nonumber \\
		&\times \overline{x}[[k-k_p+q]_N,[l-l_{p}]_M] + \overline{w}[k, l], \forall k,l,\label{eq:effective_dd}
	\end{align}
	where $\overline{w}[k,l]\sim\mathcal{CN}(0,\sigma_n^2)$ denotes the additive noise with variance $\sigma^{2}_{n}$, and $\alpha_p(k,l,q)$ is defined as
	\begin{equation}
		\begin{split}
			&\alpha_p(k,l,q)  \triangleq \\
			&\begin{cases}
				\beta_p(q),& l \in \{l_p,  \cdots , M -  1\} \\
				\left(\beta_p(q)-1\right)  e^{-j2\pi \frac{[k  -  k_{p} + q]_N}{N}},  &l \in \{0, \cdots ,l_p- 1\}
			\end{cases} ,
		\end{split}
	\end{equation}
	with $\beta_p(q)\triangleq \lim_{q^{\prime}\rightarrow q}\frac{e^{-j2\pi (-q^{\prime }-\kappa_p)}-1}{e^{-j\frac{2\pi}{N}(-q^{\prime}-\kappa_p)}-1}$. Note that without the fractional Doppler shift, i.e., $\kappa_{p}=0$, $\frac{1}{N} |\alpha_p(k,l,0)| \approx 1$ and $\frac{1}{N} |\alpha_p(k,l,q)|\approx0, \forall q\neq 0$ as $N$ is typically large. Thus, the terms in \eqref{eq:effective_dd} associated with $\forall q\neq0$ are regarded as IDI. 
	
	Define $\overline{\mathbf{x}} \triangleq \text{vec}\left(\{\overline{x}\left[k,l\right]\}\right)$, $\overline{\mathbf{y}}\triangleq\text{vec}\left(\{\overline{y}\left[k,l\right]\}\right)$, $\overline{\mathbf{w}} \triangleq \text{vec}\left(\{\overline{w}[k, l]\}\right)$, and $\mathbf{H}_{\text{eff}}\in \mathbb{C}^{MN\times MN}$ as the effective channel matrix with the $(\left(kM+l\right),\left(k^{\prime}M+l^{\prime}\right))$-th entry summarizing the channel coefficient between $\overline{y}[k, l]$ and $\overline{x}[k^{\prime}, l^{\prime}]$, thus yielding $\mathbf{\overline{y}}=\mathbf{H}_{\text{eff}}\mathbf{\overline{x}} + \overline{\mathbf{w}}$. Assuming perfect channel knowledge at the receiver, the maximum a posteriori (MAP) detector for the transmitted symbols $\overline{\mathbf{x}}$ is given below:
	\begin{equation}
		\overline{\mathbf{x}}_{\text{MAP}} = \operatorname*{argmax}_{\{\overline{x}\left[k, l\right]\} \in \mathcal{Q}^{MN}} p \left(\overline{\mathbf{x}}|\overline{\mathbf{y}}, \mathbf{H}_{\text{eff}}\right), \label{eq:origin_MAP}
	\end{equation}
	which has exponential computational complexity with respect to $MN$, preventing its practical use. Although the GNN-based detector~\cite{Xufan_gcn} is of low complexity, it fails to efficiently exploit prior information of transmitted symbols.
	In the following, we develop an AMP-GNN-based detector, which mitigates major drawbacks of the GNN-based detector.
	\begin{figure}[t!]
		\centering
		\includegraphics[width=0.95\linewidth]{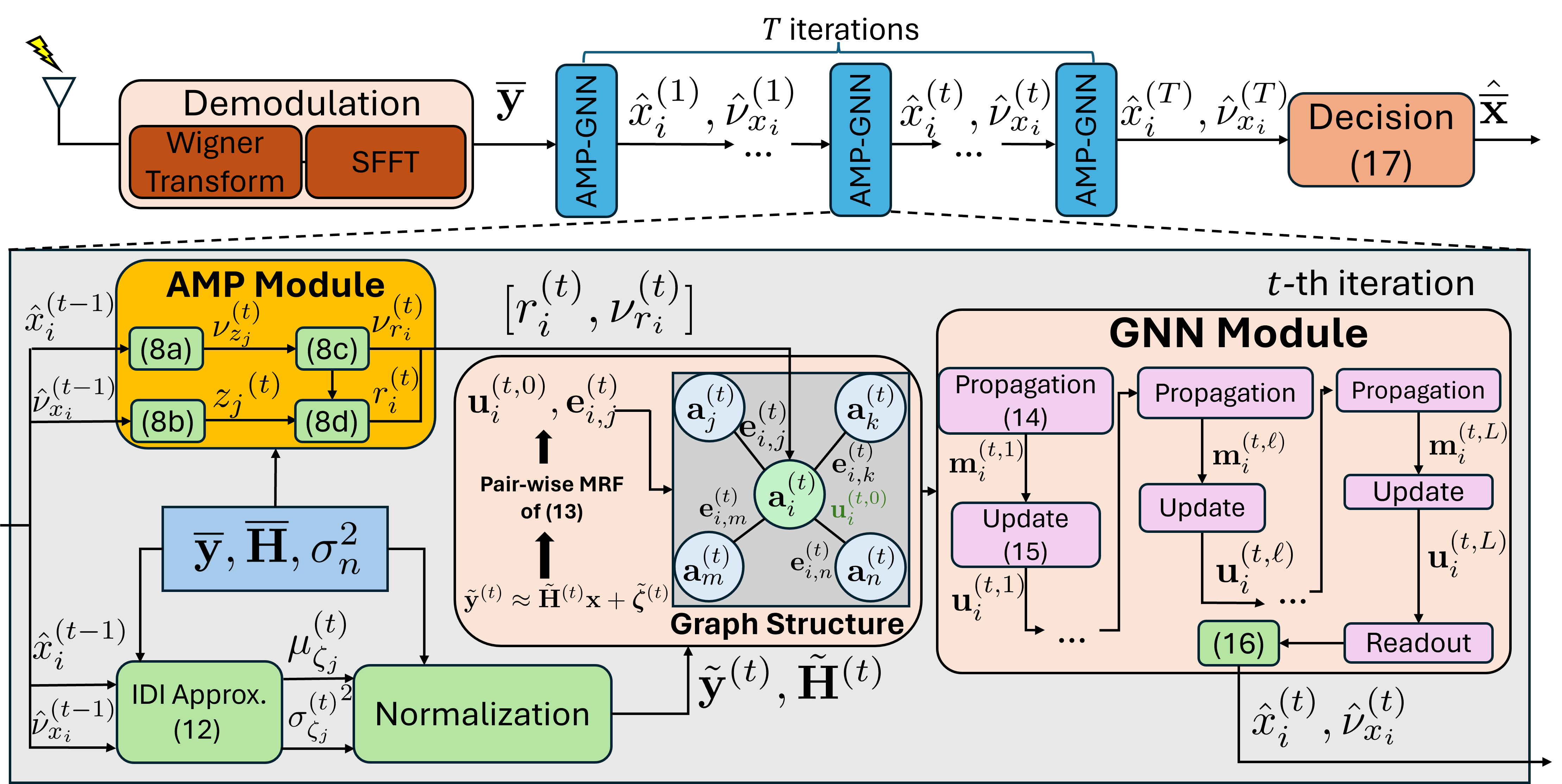}
		\caption{.~~AMP-GNN-based detector for OTFS systems.}
		\label{fig:amp_gnn_architecture}
		\vspace{-0.5cm}
	\end{figure}

	\section{AMP-GNN-based detector for OTFS Systems}\label{sec:AMP-GNN-Detector}
	In this section, we develop an AMP-GNN-based detector for OTFS systems, which alternates between an AMP module and a GNN module over $T$ iterations as depicted in Fig.~\ref{fig:amp_gnn_architecture}. A real-valued approximate signal model is first developed for ease of implementation, followed by operations of the AMP module and the GNN module. To further reduce the detection overhead, an IDI approximation scheme is also proposed. 
	\subsection{Real-valued Approximate Signal Model}
	As shown in \eqref{eq:effective_dd}, the number of non-zero elements in each row of $\mathbf{H}_{\text{eff}}$ is $PN$, which results in a complex pair-wise MRF. By exploiting the property that $|\alpha_p(k,l,q)|$ peaks at $q=0$ and decreases as $\lvert q \rvert$ increases, only $2N_{o}+1$ ($2N_{o}\ll N$) most significant IDI symbols in \eqref{eq:effective_dd} are retained \cite{Raviteja2018}, leading to the following approximation:
	\begin{align}
		\overline{y}[k,l] \approx &\overline{y}_{\text{a}}\left[k,l\right] \triangleq  \frac{1}{N} \sum_{p=1}^{P} \overline{h}_p e^{j2\pi \frac{(l-l_{p})(k_p+\kappa_p)}{MN}} \sum_{q=-N_o}^{N_o} \alpha_p(k,l,q)  \nonumber \\
		&\times  \overline{x}[[k-k_{p}+q]_N,[l-l_{p}]_M] + \overline{w}[k, l],\forall k,l.
		\label{eq:input-output}
	\end{align}
	The vector form of \eqref{eq:input-output} is expressed as follows:
	\begin{align}
		\overline{\mathbf{y}}\approx \overline{\mathbf{y}}_{\text{a}} = \overline{\mathbf{H}} \overline{\mathbf{x}} + \overline{\mathbf{w}},
		\label{eq:vec_input_output}
	\end{align}
	where $\overline{\mathbf{y}}_{\text{a}}\triangleq \text{vec} \left(\{\overline{y}_{\text{a}}\left[k,l\right]\}\right)$, and the entry at the $j=(kM+l)$-th row and $i=(k^{\prime}M+l^{\prime})$-th column of $\overline{\mathbf{H}}$ is given as $\overline{h}_{ji}\triangleq\frac{1}{N}\sum_{p=1}^{P}\sum_{q=-N_o}^{N_{o}}\overline{h}_p e^{j2\pi \frac{(l-l_{p})(k_{p}+\kappa_{p})}{MN}} \alpha_p(k,l,q)\delta(k^{\prime}-[k-k_p+$ $q]_N)\delta(l^{\prime}-[l-l_p]_M)$ that can be extracted from ${\mathbf{H}}_{\text{eff}}$. 
	Since $\overline{\mathbf{H}}$ is sparser than $\mathbf{H}_{\text{eff}}$, the detection complexity can be reduced as the insignificant interfering symbols are ignored.

	Since real-valued GNNs are more convenient to implement, $\overline{\mathbf{x}}$, $\overline{\mathbf{y}}$, $\overline{\mathbf{H}}$ are transformed to their real-valued representations, i.e., 
	${\mathbf{x}}\triangleq\left[\mathfrak{R} (\overline{\mathbf{x}})^{\mathrm{T}}, \mathfrak{I}(\overline{\mathbf{x}})^{\mathrm{T}}\right]^{\mathrm{T}}\in \mathbb{R}^{2MN\times 1}$,
	${\mathbf{y}}\triangleq\left[{\mathfrak{R}}(\overline{\mathbf{y}})^{\mathrm{T}}, \mathfrak{I}(\overline{\mathbf{y}})^{\mathrm{T}}\right]^{\mathrm{T}}\in \mathbb{R}^{2MN\times1}$,
	$\mathbf{H} \triangleq \begin{bmatrix} \mathfrak{R} (\overline{\mathbf{H}}) & -\mathfrak{I}(\overline{\mathbf{H}}) \\ \mathfrak{I}(\overline{\mathbf{H}}) & \mathfrak{R} (\overline{\mathbf{H}}) \end{bmatrix} \in \mathbb{R}^{2MN\times2MN}$, 
	where $\mathfrak{R}(\cdot)$ and $\mathfrak{I}(\cdot)$ return the real and imaginary part of complex-valued input, respectively. Thus, we yield $\mathbf{y} \approx \mathbf{H} \mathbf{x} + \mathbf{w}$,
	where $\mathbf{w} \triangleq [w_{0},\cdots, w_{2MN-1}]^{\rm{T}} \sim \mathcal{N}(\bm{0}, \frac{\sigma_n^2}{2}\mathbf{I})$ with $\mathbf{I}$ denoting the identity matrix. We also define $\mathcal{S}\triangleq \{0,1,\cdots, 2MN-1\}$. 
	The column and row index set of the non-zero elements at the $j$-th row and $i$-th column of $\overline{\mathbf{H}}$, and the index set of IDI symbols are respectively defined as follows:
	\begin{subequations}
		\begin{align}
			{\mathcal{I}}(j) &\triangleq \{i\!+\!a |i=[\lfloor [j]_{MN} \slash M \rfloor \!- \!k_{p}\!- \!q]_N \cdot M + [[j]_M \! - \!l_{p}]_M, \nonumber \\
			&  q \in \{-N_o,\cdots, N_o\},  p \in \{1, \cdots, P\}, a=0,MN\},\label{eq:set0} \\
			\mathcal{L}(i) &\triangleq \{j \! + \!a |j=[\lfloor [i]_{MN} \slash M \rfloor \! + \! k_{p}\!+ \!q]_N \cdot M + [[i]_M \! + \! l_{p}]_M, \nonumber \\
			&  q \in \{-N_o,\cdots, N_o\},  p \in \{1, \cdots, P\}, a=0,MN\}, \label{eq:set1} \\
			\tilde{\mathcal{I}}(j) &\triangleq \{i\!+\! a |i=[\lfloor [j]_{MN} \slash M \rfloor \! - \! k_p \!-\!q]_N \cdot M + [[j]_M-l_p]_M, \nonumber \\
			& q \in \{-N_o,\cdots, N_o\}\setminus \{0\}, p \in \{1, \cdots, P\}, a=0,MN\}. \label{eq:set2}
		\end{align}
	\end{subequations}
	Based on the real-valued approximate signal model, the AMP module and GNN module are executed sequentially in the $t$-th iteration of the proposed detector, as detailed in the sequel.

	\subsection{AMP Module}
	Denote $\hat{x}_i^{(t)}$ and $\hat{\nu}_{x_i}^{(t)}$ as the estimates for the mean and variance of $x_i$, i.e., the $i$-th element of $\mathbf{x}$, obtained by the GNN module in the $t$-th iteration, respectively. Define $y_j$ and $z_j$ as the $j$-th element of  $\mathbf{y}$ and $\mathbf{z}\triangleq \mathbf{H}\mathbf{x}$, respectively, and $h_{ji}$ as the $(j,i)$-th entry of $\mathbf{H}$. We initialize $\hat{x}_i^{(0)}=0$, $\hat{\nu}_{x_i}^{(0)}=0.5$, $z_j^{(0)}=y_j$, $\nu_{z_j}^{(0)}=\sum_{i\in \mathcal{I}(j)} h_{ji}^2 \hat{\nu}_{x_i}^{(0)}, \forall i, j$. The AMP module decomposes the estimation of $\mathbf{x}$ to $2MN$ scalar estimation problems and operates in the $t$-th iteration as follows:
	\begin{subequations}
		\label{eq:AMP}
		\begin{align}
			\nu_{z_j}^{(t)} &= \sum\nolimits_{i\in \mathcal{I}(j)} h_{ji}^2 \hat{\nu}_{x_i}^{(t-1)},  j\in\mathcal{S}, \label{eq:var_z} \\ 
			z_j^{(t)} &= \sum_{i\in \mathcal{I}(j)} h_{ji} \hat{x}_i^{(t-1)}-\frac{\nu_{z_j}^{(t)} (y_j-z_j^{(t-1)})}{\nu_{z_j}^{(t-1)}+ \frac{1}{2} \sigma_n^2}, j\in\mathcal{S},  \label{eq:mean_z} \\
			\nu_{r_i}^{(t)} &= \left(\sum\nolimits_{j \in \mathcal{L}(i)} \frac{\lvert h_{ij} \rvert^2}{\nu_{z_j}^{(t)} + \frac{1}{2}\sigma_n^2}\right)^{-1},  i \in\mathcal{S}, \label{eq:var_r} \\ 
			r_{i}^{(t)} &=  \hat{x}_i^{(t-1)}+\nu_{r_i}^{(t)} \sum\nolimits_{j \in \mathcal{L}(i)} \frac{h_{ij} (y_j-z_j^{(t)})}{\nu_{z_j}^{(t)}+ \frac{1}{2}\sigma_n^2}, i\in\mathcal{S}, \label{eq:mean_r}
		\end{align}
	\end{subequations}
	where $z_j$ is approximated by a Gaussian random variable with variance $\nu_{z_j}^{(t)}$ and mean $z_j^{(t)}$ obtained respectively in \eqref{eq:var_z} and \eqref{eq:mean_z}. Given the residuals $\{y_j-z_j^{(t)}\}$'s, posterior distributions of $\{x_i\}$'s are also Gaussian with variances and means estimated in \eqref{eq:var_r} and \eqref{eq:mean_r}, respectively. These values are passed to the GNN module as prior information. 

	\subsection{GNN Module}
	In this subsection, the pair-wise MRF, which is the core of the GNN module, is first introduced. An IDI approximation scheme is then developed to simplify the pair-wise MRF for complexity reduction. Finally, we present the GNN processing steps that derive posterior estimates of transmitted symbols.
	\subsubsection{\textbf{Pair-wise MRF}}
	\label{sec3A}
	Pair-wise MRF is a probabilistic graphical model that characterizes the joint distribution of random variables. It is modeled via an undirected graph, where each node represents a random variable, and an edge connecting two nodes captures their dependencies. 
	For OTFS data detection, the random variables correspond to the transmitted symbols $\mathbf{x}=[x_0, \cdots, x_{2MN-1}]^{\mathrm{T}}$. The joint posterior distribution characterized by the pair-wise MRF is factorized as follows~\cite{scotti2020graph}:
	\begin{equation}
		p({\mathbf{x}}|{\mathbf{y}}, {\mathbf{H}}) \approx \frac{1}{Z} \prod_{i\in \mathcal{S}} \phi ({x}_i; {\mathbf{y}}, {\mathbf{H}}) \prod_{\substack{j \in {\mathcal{N}}(i)}}\psi ({x}_i, {x}_j; {\mathbf{H}}) \label{eq:posteriori},
	\end{equation}
	where $Z$ is a normalizing factor and $\mathcal{N}(i) \triangleq \{j | {\mathbf{h}}_{i}^{\text{T}}{\mathbf{h}}_{j} \neq 0\}$. Let $p(x_i)$ be the prior distribution of $x_i$. The self-potential function of ${x}_i$, denoted as $\phi ({x}_i; {\mathbf{y}}, {\mathbf{H}})$, and the pair-potential function of ${x}_i$ and ${x}_j$, denoted as $\psi ({x}_i, {x}_j; {\mathbf{H}})$, are respectively expressed as follows:
	\begin{subequations}
		\label{eq:pair_potential}
		\begin{align}
			\phi ({x}_i; {\mathbf{y}}, {\mathbf{H}}) &= e^{\frac{2}{\sigma_n^2}({\mathbf{y}}^{\mathrm{T}}{\mathbf{h}}_i {x}_i - \frac{1}{2} {\mathbf{h}}_i^{\mathrm{T}} {\mathbf{h}}_i {x}_i^2)} p({x}_i), \\
			\psi ({x}_i, {x}_j; {\mathbf{H}}) &= e^{-\frac{2}{\sigma_n^2}{\mathbf{h}}_i^{\mathrm{T}} {\mathbf{h}}_j {x}_i {x}_j}. 
		\end{align}
	\end{subequations}
	
	When GNN is applied to learn the posterior distributions of random variables from a pair-wise MRF, each node is embedded with a feature vector that is iteratively updated by exchanging messages with neighboring nodes, in the form of node features and edge attributes. Each node aggregates and encodes the received messages. 
	After several iterations, the posterior distribution is read out from the updated feature vector with a multi-layer perceptron (MLP). 
	However, the significant number of node-pairs in the pair-wise MRF due to the IDI leads to high computational complexity of message aggregation. To achieve cost-effective data detection, we next approximate the IDI via learned Gaussian distributions.
	
	\subsubsection{\textbf{IDI Approximation}}
	\label{sec:idi}
	By exploiting the index sets $\mathcal{I}(j)$ and $\tilde{\mathcal{I}}(j), j\in \mathcal{S}$, the $j$-th element of ${\mathbf{y}}$ is given as follows:
	\begin{align}
		{y}_{j} \approx \sum_{i \in {{\mathcal{I}}(j)}\setminus{\tilde{\mathcal{I}}(j)}} {h}_{ji}{x}_i + \sum_{i^{\prime} \in \tilde{\mathcal{I}}(j)} {h}_{ji^{\prime}} {x}_{i^{\prime}} + {w}_j.
		\label{eq:simple_y}
	\end{align}
	Since the IDI is caused by the power leakage of transmitted symbols due to insufficient Doppler resolution, ${x}_{i^{\prime}},{i^{\prime}\in\mathcal{\tilde{I}}\left(j\right)}$ has much less impact on ${y}_{j}$ than other symbols \cite{power_leakage}. Inspired by the hybrid data detector in \cite{hybrid_mappic} that differentiates OTFS symbols according to their path gains, we approximate the IDI and noise terms in \eqref{eq:simple_y}, i.e. $\sum_{i^{\prime} \in \tilde{\mathcal{I}}(j)} {h}_{ji^{\prime}} {x}_{i^{\prime}} + {w}_j$, by ${\zeta}_j\sim \mathcal{CN}({\mu}_{{\zeta}_j},\sigma^2_{{\zeta}_j})$ to derive a simplified pair-wise MRF. 

	In each iteration, ${\mu}_{{\zeta}_j}$ and $\sigma^2_{{\zeta}_j}$ are estimated by the proposed AMP-GNN-based detector via learned estimates for the mean and variance of $x_i$ in the previous iteration, which is given by 
	\begin{subequations}
		\label{eq:update_if}
		\vspace{-1em}
		\begin{align}
			\mu_{\zeta_j}^{(t)} &= \sum\nolimits_{i^{\prime} \in \tilde{\mathcal{I}}(j)} h_{ji^{\prime}} \hat{x}_{i^{\prime}}^{(t-1)}, j\in\mathcal{S}, \label{eq:update_a} \\
			{\sigma^{(t)}_{\zeta_j}}^{2} &= \sum\nolimits_{i^{\prime} \in \tilde{\mathcal{I}}(j)} h_{ji^{\prime}}^2 \hat{\nu}_{x_{i^{\prime}}}^{(t-1)} + \frac{\sigma^2_n}{2},j\in\mathcal{S}. \label{eq:update_b}
		\end{align}
	\end{subequations}
	The normalized received signal $\tilde{\mathbf{y}}^{(t)}\triangleq [\tilde{y}_{0}^{(t)},\cdots, \tilde{y}_{2MN-1}^{(t)}]^{\mathrm{T}}$ with $\tilde{y}_{j}^{(t)} \triangleq (y_{j}-\mu_{\zeta_{j}}^{(t)})\slash \sigma_{\zeta_{j}}^{(t)}$ is approximated as follows:
	\begin{align}
		\tilde{\mathbf{y}}^{(t)} & \approx \tilde{\mathbf{H}}^{(t)} \mathbf{x}+ \tilde{\bm{\zeta}}^{(t)},  \label{eq:new_vec}
	\end{align}
	where $\tilde{\bm{\zeta}}^{(t)}\sim \mathcal{N}(\mathbf{0}, \mathbf{I})$ and the $(j,i)$-th entry of $\tilde{\mathbf{H}}^{\left(t\right)}$ is denoted as $\tilde{h}_{ji}^{\left(t\right)} \triangleq h_{ji} \slash \sigma_{\zeta_j}^{(t)}$. By replacing $\mathbf{y}$, $\mathbf{H}$, $\mathcal{N}(i)$ in \eqref{eq:posteriori} and \eqref{eq:pair_potential} with $\tilde{\mathbf{y}}^{(t)}$, $\tilde{\mathbf{H}}^{(t)}$, $\tilde{\mathcal{N}}^{(t)}(i)\!\triangleq \! \{j | \tilde{\mathbf{h}}_i^{(t)^{\mathrm{T}}}\tilde{\mathbf{h}}_j^{(t)}\! \neq\! 0 \}$, we yield the pair-wise MRF of \eqref{eq:new_vec}, which is simplified as the number of node-pairs $\sum_i \lvert \tilde{\mathcal{N}}^{(t)}(i) \rvert \! \ll \! \sum_i \lvert \mathcal{N}(i) \rvert $. Note that this is different from the AMP-GNN network in~\cite{Hengtao_GNN-MIMO} with unchanged pair-wise MRF across iterations.
	\subsubsection{\textbf{GNN Processing}}
	To leverage prior information obtained from the AMP module, the node attributes are embedded as $\mathbf{a}_i^{(t)}\triangleq [r_i^{(t)}, \nu_{r_i}^{(t)}], \forall i$, which augments the GNN structure in \cite{Xufan_gcn} without considering node attributes. The $N_u$-dimensional feature vector for node $i$ is initialized as $\mathbf{u}_i^{(t, 0)}=\mathbf{W}_1[\tilde{\mathbf{y}}^{(t)^{\mathrm{T}}} \tilde{\mathbf{h}}_i^{(t)}, \tilde{\mathbf{h}}_i^{(t)^{\mathrm{T}}} \tilde{\mathbf{h}}_i^{(t)}]^{\mathrm{T}}+\mathbf{b}_1, i \in\mathcal{S}$, where $\mathbf{W}_1 \in \mathbb{R}^{N_u\times 2}$, $\mathbf{b}_1 \in  \mathbb{R}^{N_u \times 1}$ are learnable and $N_u$ is a hyperparameter. The edge attributes are denoted as $\mathbf{e}_{i, j}^{(t)}= {\tilde{\mathbf{h}}_i^{(t)^{\mathrm{T}}}} {\tilde{\mathbf{h}}_j^{(t)}},\forall i,j$. 
	 	
	The node feature vectors are updated through $L$ rounds of graph convolution, each of which consists of a \textbf{Propagation Step} and an \textbf{Update Step} \cite{scotti2020graph}. After $L$ graph convolution rounds, a \textbf{Readout Step} is implemented to derive the posterior estimates of $\mathbf{x}$. These steps are detailed below:
	\begin{itemize}
		\item \textbf{Propagation Step}: In the $\ell$-th graph convolution round, the message from node $x_j$ to node $x_i$ is expressed as $\mathbf{m}_{j \rightarrow i}^{(t, \ell)}=[\mathbf{u}_{i}^{(t, \ell-1)}, \mathbf{u}_{j}^{(t, \ell-1)},\mathbf{e}_{i,j}^{(t)}]$. Each node aggregates the incoming messages from all its neighboring nodes and encodes the aggregated message $\mathbf{m}_i^{(t, \ell)}\in \mathbb{R}^{N_u\times 1}$ through an MLP $f_{\bm{\theta}}\left(\cdot\right)$ with parameters $\bm{\theta}$ as follows: 
		\begin{align}
			\mathbf{m}_i^{(t, \ell)} &= f_{\bm{\theta}} \left(\sum\nolimits_{j\in {\tilde{\mathcal{N}}^{(t)}}(i)}\mathbf{m}_{j \rightarrow i}^{(t, \ell)}\right).  \label{eq:sum_mlp}
		\end{align}
		\item \textbf{Update Step}: Each node passes the encoded message and its attribute to a gated recurrent unit (GRU) $f_{\bm{\varphi}}(\cdot)$ with parameters $\bm{\varphi}$ to update its node feature vector as follows:
		\begin{subequations}
			\label{eq:update_step}
			\vspace{-1em}
			\begin{align}
				\mathbf{s}_i^{(t, \ell)} &= f_{\bm{\varphi}}\left(\mathbf{s}_i^{(t, \ell-1)}, \left[\mathbf{m}_i^{(t, \ell)}, \mathbf{a}_i^{(t)}\right] \right), \label{eq:update_step_1}\\
				\mathbf{u}_i^{(t, \ell)} &= \mathbf{W}_2 \cdot \mathbf{s}_i^{(t, \ell)} + \mathbf{b}_2, \label{eq:update_step_2}
			\end{align}
		\end{subequations}
		where $\mathbf{s}_i^{(t, \ell)} \in \mathbb{R}^{N_{h}\times 1}$ denotes the GRU hidden state, $\bm{\varphi}$, $\mathbf{W}_2 \in \mathbb{R}^{N_u \times N_{h}}$, $\mathbf{b}_2 \in \mathbb{R}^{N_u\times 1}$ are learnable, and $N_{h}$ is a hyperparameter. 
		\item \textbf{Readout Step}: Node feature vector $\textbf{u}_i^{(t, L)}$ is passed to MLP $f_{\bm{\omega}}\left(\cdot\right)$ with a $|\mathcal{Q}_{\text{R}}|$-dimensional Softmax output layer to update the posterior distribution of $x_i,i\in\mathcal{S}$ as $[p^{(t)}_{\text{AMP-GNN}}(x_i=s|\textbf{y},\textbf{H})]_{s\in \mathcal{Q}_\text{R}} = f_{\bm{\omega}}(\mathbf{u}_i^{(t, L)})$, where $\mathcal{Q}_{\text{R}} \triangleq \{\mathfrak{R}(s)| s \in \mathcal{Q} \}$ and $\bm{\omega}$ is learnable. The posterior mean and variance of $x_i$ are estimated as follows:
		\begin{subequations}
			\label{eq:new_xnu}
			\begin{align}
				\hat{x}_i^{(t)} &= \sum_{s\in \mathcal{Q}_{\text{R}}} s \cdot p^{(t)}_{\text{AMP-GNN}}(x_i=s|\textbf{y},\textbf{H}), \\
				\hat{\nu}_{x_i}^{(t)} &= \sum_{s\in \mathcal{Q}_{\text{R}}} (s-\hat{x}_i^{(t)})^2 \cdot p^{(t)}_{\text{AMP-GNN}}(x_i=s|\textbf{y},\textbf{H}).
			\end{align}
		\end{subequations}
	\end{itemize}
	
	Since $\mathbf{H}$ deviates from the i.i.d. sub-Gaussian assumption \cite{Zhengdao_UAMP}, $\{\hat{x}_i^{(t)}\}$'s and $\{\hat{\nu}_{x_i}^{(t)}\}$'s are then passed to the AMP module to improve its symbol estimates in the next iteration. After $T$ iterations, the transmitted symbols are detected as follows:
	\begin{align}
		\hat{\overline{x}}_i = \operatorname*{argmin}_{s \in \mathcal{Q}} {\Vert \hat{\overline{x}}^{(T)}_i-s\Vert}_2 , \quad i\in\{0,\cdots,MN-1\},  
		\label{eq:deci}
	\end{align}
	where $\hat{\overline{x}}^{(T)}_i = \hat{x}_{i}^{\left(T\right)} + j \cdot \hat{x}_{i+MN}^{\left(T\right)}$. 
	Denote $\hat{\overline{\mathbf{x}}} \triangleq [\hat{\overline{x}}_{0}^{(T)},$ $\cdots, \hat{\overline{x}}_{MN-1}^{(T)}]^{\mathrm{T}} \triangleq f_{\text{AMP-GNN}}(\mathbf{\overline{y}}, \mathbf{H}_{\text{eff}}, \sigma_n^2; \bm{\Lambda})$, where $\bm{\Lambda} \triangleq \{\mathbf{W}_1, \mathbf{b}_1, \mathbf{W}_2, \mathbf{b}_2, \bm{\theta}, \bm{\varphi}, \bm{\omega}\}$ encapsulates all learnable parameters of the proposed AMP-GNN-based detector. 
		
	\vspace{-0.5em}
	\subsection{Training Method and Complexity Analysis}
	\label{sec:train_method}
	A training dataset $\mathcal{D}$ is randomly generated, where each sample $d$ consists of the transmitted symbols $\overline{\mathbf{x}}_d$, received signal $\overline{\mathbf{y}}_d$, and effective channel matrix ${\mathbf{H}}_{\text{eff},d}$. The transmitted symbols are uniformly sampled from $\mathcal{Q}$ and distorted by a time-varying channel with $l_p$ and $k_p$ randomly sampled from $\{0, \cdots, l_{\text{max}}\}$ and $\{-k_{\text{max}}, \cdots, k_{\text{max}}\}$ respectively at noise variance $\sigma_{\text{train}}^2$. 
	Thus, ${\mathbf{H}}_{\text{eff},d}$ and $\overline{\mathbf{y}}_d$ can be obtained from \eqref{eq:effective_dd}. We optimize $\bm{\Lambda}$ by minimizing the following $l_2$-loss function:
	\begin{align}
		\frac{1}{|\mathcal{D}|} \sum_{\overline{\mathbf{x}}_d \in \mathcal{D}} \Vert \overline{\mathbf{x}}_d - f_{\text{AMP-GNN}}(\mathbf{\overline{y}}_d, \mathbf{{H}}_{\text{eff},d}, \sigma_{\text{train}}^2;\bm{\Lambda}) \Vert^2_2.
	\end{align}

	In each iteration, the computational complexity of the AMP module is given as $\mathcal{O}(16MN(2N_o+1)P)$~\cite{Zhengdao_UAMP}, while that of the GNN module is dominated by the graph convolutions. With the proposed IDI approximation scheme, the complexity of message aggregation in \eqref{eq:sum_mlp} is $\mathcal{O}(4MN P^2N_u)$, which is much lower than $\mathcal{O}(8MN(2N_o+1)P^2N_u)$ of the GNN-based detector~\cite{Xufan_gcn}\footnote{Detailed computational complexity analysis is available in Appendix \ref{sec:FLOPS_Count}.}. The computational complexity of $f_{\bm{\theta}}(\cdot)$ in \eqref{eq:sum_mlp} depends on implementation and that of \eqref{eq:update_step} is dominated by the GRU operations given as $\mathcal{O}(6MN(N_h N_u+N_h^2))$. When $f_{\bm{\theta}}(\cdot)$ is a three-layer MLP, the computational complexity of the AMP-GNN-based detector scales as $\mathcal{O}(16MN(2N_o+1)P+ 4MN P^2 N_u L + 2MN[(2N_u+1)N_{h_1} + N_{h_1}N_{h_2}+N_{h_2}N_u]L + 6MN(N_h N_u+N_h^2)L)$ with $N_{h_1}$ and $N_{h_2}$ denoting the hidden layer dimensions.

	\vspace{-0.2em}
	\section{Simulation Results}\label{sec:simulations}
	We evaluate the proposed AMP-GNN-based detector by simulating an OTFS system with $\Delta f = 1\slash \Delta T= 30$ kHz, $M = 64$, and $N=16$.
	We set $l_{\text{max}}=8$, $k_{\text{max}}=2$, $N_o = 5$, and $\overline{h}_p \sim \mathcal{CN}(0, \sigma_p^2) $ with $\sigma_p^2=e^{-0.1l_p} \slash  \sum_{i=1}^{P} e^{-0.1l_i}$~\cite{hybrid_mappic}. Besides, $T$, $L$, $N_{u}$, and $N_{h}$ are configured as $15$, $2$, $8$, and $12$, respectively. Both $f_{\bm{\theta}}\left(\cdot\right)$ and $f_{\bm{\psi}}\left(\cdot\right)$ are three-layer MLPs with hidden layer dimensions as $N_{h_1}=16$ and $N_{h_2}=12$. For comparison purposes, we simulate three baselines, including the MP-based \cite{Raviteja2018}, AMP-based \cite{amp_otfs}, and GNN-based \cite{Xufan_gcn} detectors. Two variants of the AMP-GNN-based detectors, namely AMP-GNN-v1 and AMP-GNN-v2, are also examined to validate the benefits of the proposed IDI approximation scheme. Specifically, AMP-GNN-v1 ignores the IDI by setting $\hat{x}_{i}^{(t)}=0$ and $\hat{\nu}_{x_{i}}^{(t)}=0, i\in\tilde{\mathcal{I}}\left(j\right)$, while AMP-GNN-v2 does not implement IDI approximation. The results are averaged over 50,000 independent trials, and all experiments were conducted on a server with an Intel Xeon 4210R CPU, an NVIDIA RTX 3090 GPU, and 128GB RAM.
	\begin{figure}[t!]
		\centering
		\begin{subfigure}{0.21\textwidth}
			\centering
			\includegraphics[width=\linewidth]{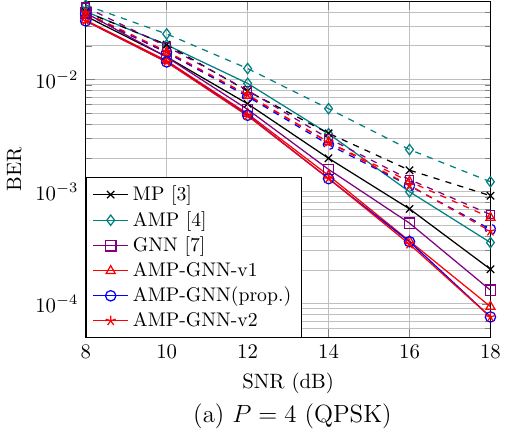}
		\end{subfigure}
		\begin{subfigure}{0.21\textwidth}
			\centering
			\includegraphics[width=\linewidth]{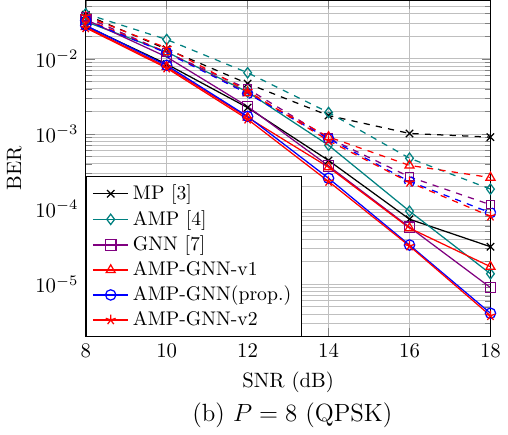}
		\end{subfigure}
		\begin{subfigure}{0.21\textwidth}
			\centering
			\includegraphics[width=\linewidth]{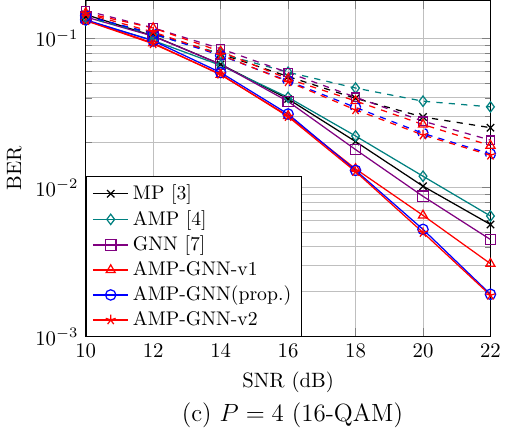}
		\end{subfigure}
		\begin{subfigure}{0.21\textwidth}
			\centering
			\includegraphics[width=\linewidth]{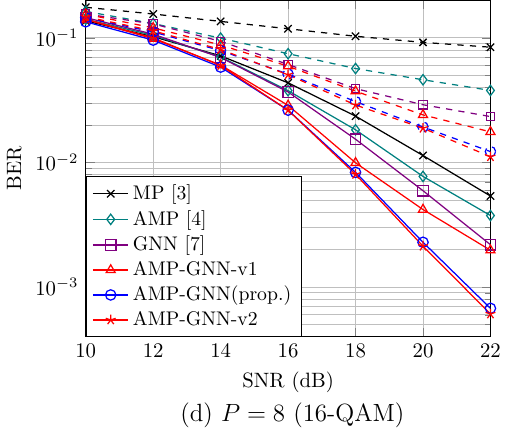}
		\end{subfigure}
		\caption{.~~BER vs. SNR. The solid and dashed curves represent cases with perfect ($\sigma_e^2=0$) and imperfect CSI ($\sigma_e^2=-20$ dB), respectively, where $\overline{h}_{p}$ is contaminated by additive Gaussian noise $n_{e,p} \sim \mathcal{CN}(0, \frac{1}{P}\sigma_e^2)$.}
		\label{fig:16qam}
		\vspace{-0.5cm}
	\end{figure}

	The BER performance achieved by different OTFS data detectors versus the SNR with quadrature phase shift keying (QPSK) and 16-QAM ($\sigma_{\text{train}}^2=-15$ and $-20$ dB respectively) is shown in Fig. \ref{fig:16qam}. The AMP-based detector behaves far from optimal and the MP-based detector suffers from severe performance degradation when $P\!=\!8$. In line with \cite{Xufan_gcn}, the GNN-based detector outperforms the MP-based and AMP-based detectors, while the proposed AMP-GNN-based detector and its variants secure further and significant BER reductions. The latter is attributed to the prior information obtained from the AMP module that helps improve the estimates of the GNN module. Besides, compared with the AMP-GNN-based detector, the BER achieved by AMP-GNN-v1 notably increases at the high-SNR regime, especially with $P\!=\!8$ or 16-QAM. In these scenarios, the IDI has large impact and simply ignoring them compromises the estimation accuracy of the GNN module. Moreover, the AMP-GNN-based detector performs indistinguishably with AMP-GNN-v2 despite with the proposed IDI approximation scheme. Besides, the proposed AMP-GNN-based detector guarantees performance gains over baselines even with imperfect channel state information (CSI).

	Table \ref{tab:flops_time} compares the complexity of different OTFS  detectors in terms of the required number of floating-point operations (FLOPs)\footnote{The required number of FLOPs is counted by the fvcore library available at: \url{https://github.com/facebookresearch/fvcore}, while those of the unsupported operations, e.g. sparse matrix operations, are counted manually.} and runtime. It is observed that the AMP-based detector has the lowest complexity but worst BER performance. Thanks to the aid of AMP, AMP-GNN-v2 achieves pronounced BER improvements over the GNN-based detector. With IDI approximation, the AMP-GNN-based detector shows much lower complexity than that of AMP-GNN-v2 at negligible loss of BER performance, which demonstrates its benefits of cost-effective OTFS data detection. It is noteworthy that when $P=4$, the GNN-based detector incurs shorter runtime than the MP-based detector despite requiring slightly more FLOPs owing to the GPU acceleration for neural networks.
	\begin{table}[t]
		\renewcommand{\arraystretch}{1.0}
		\begin{tabular}{l|c|c|c|c|c|c}
			\hline
			\multirow{2}{*}{Detector} & \multicolumn{3}{c|}{$P=4$} & \multicolumn{3}{c}{$P=8$} \\
			\cline{2-7}
			& FLOPs& Time & BER & FLOPs & Time & BER \\
			\hline\hline
			MP-based \cite{Raviteja2018}             & $20.01$   & $100.0$ &10.2 &$40.01$ & $120.4$ & 11.4 \\ \hline
			AMP-based \cite{amp_otfs}            & $1.23$    & $11.3$  &11.9 &  $2.32$ & $12.7$ & 7.7  \\ \hline
			GNN-based \cite{Xufan_gcn}            & $18.44$   & $26.1$ &8.7  &$25.56$ & $36.5$ & 5.9 \\ \hline
			AMP-GNN         & $10.00$   & $26.8$ &5.1  &$14.84$ & $28.1$  & 2.2 \\ \hline
			AMP-GNN-v1      & $10.00$   & $25.9$ &6.5  &$14.84$ & $27.5$  & 4.2 \\ \hline
			AMP-GNN-v2      & $27.93$   & $37.7$ &5.0  &$45.03$ & $46.6$ & 2.1 \\ \hline
		\end{tabular}
		\caption{.~~FLOPs ($\times 10^{7}$), runtime (ms), and BER ($\times 10^{-3}$) comparisons (16-QAM, $\text{SNR}=20$ dB)\protect \footnotemark.}
		\label{tab:flops_time}
		\vspace{-0.5cm}
	\end{table}
	\footnotetext{{The FLOPs and runtime decomposition are provided in Appendix \ref{sec:FLOP_Count_Table}.}}
	\section{Conclusion}
	This letter developed an AMP-GNN-based data detector for OTFS modulation, which exploits prior information obtained from an AMP module to improve the data symbol estimates of a GNN module in an iterative manner. To reduce the computational complexity, learning-based approximation for inter-Doppler interference was further proposed. Simulation results validated the advantages of the proposed AMP-GNN-based detector over existing baselines. For future research, it will be interesting to replace AMP with more robust Bayesian optimization algorithms and extend this study for joint channel estimation-data decoding in OTFS systems. 
	\vspace{-0.1cm}
	\bibliographystyle{IEEEtran}
	\bibliography{ref}

\newpage

\appendices
\section{FLOP Counting for the AMP-GNN-based Detector}
\label{sec:FLOPS_Count}
We start with pointing out that the number of non-zero elements in each row of $\mathbf{H}$ as $2P(2N_o+1)$, which means the number of non-zero elements in $\mathbf{H}$ is $Q=2MN \times 2P(2N_o+1)$. Following the setting in Section~\ref{sec:simulations}, we have $Q=180,224$ for $P=4$ and $Q=360,448$ for $P=8$. Because the fused multiply-add (FMA) operation is commonly implemented in digital signal processing (DSP) chips, we regard each multiply-add operation as one FLOP. Besides, the required number of FLOPs for the sparse matrix-vector multiplication, i.e., $\mathbf{H}\mathbf{x}$, is $Q$. Note that our analysis excludes the computational complexity of activation functions.

\subsection{FLOP Calculation for the AMP Module}
Denote
\begin{subequations}
	\label{eq:vec_amp}
	\begin{align}
		\bm{\nu}_{\mathbf{z}}^{(t)} & \triangleq \left[\nu_{z_{0}}^{(t)},\cdots,\nu_{z_{2MN-1}}^{(t)}\right]^\mathrm{T}, \\
		\mathbf{z}^{(t)} &= \left[z^{(t)}_{0},\cdots,z^{(t)}_{2MN-1}\right]^{\mathrm{T}},\\
		\bm{\nu}_{\mathbf{r}}^{(t)} & \triangleq \left[\nu_{r_{0}}^{(t)},\cdots,\nu_{r_{2MN-1}}^{(t)}\right]^\mathrm{T},\\
		\mathbf{r}^{(t)} &= \left[r^{(t)}_{0},\cdots,r^{(t)}_{2MN-1}\right]^{\mathrm{T}}.
	\end{align}
\end{subequations}
The AMP module is implemented in its vector form given below:
\begin{subequations}
    \label{eq:vec_amp_v2}
    \begin{align}
    \bm{\nu}_{\mathbf{z}}^{(t)} &= \lvert \mathbf{H} \rvert^{\circ2} \bm{\nu}_{\mathbf{x}}^{(t-1)}, \label{eq:vec_amp0} \\
    \bm{\nu}_{\mathbf{s}}^{(t)} &= \mathbf{1} \oslash \left(\bm{\nu}_{\mathbf{z}}^{(t)}+\frac{\sigma^2_n}{2} \mathbf{1}\right), \label{eq:vec_amp1}\\
    \mathbf{s}^{(t)} &= \bm{\nu}_{\mathbf{s}}^{(t)} \odot \left(\mathbf{y} - \underbrace{\left(\mathbf{H}\hat{\mathbf{x}}^{(t-1)} - \bm{\nu}_{\mathbf{s}}^{(t)} \odot \mathbf{s}^{(t-1)} \right)}_{=\mathbf{z}^{(t)}}\right),  \label{eq:vec_amp2}\\
    \bm{\nu}_{\mathbf{r}}^{(t)} &= \mathbf{1} \oslash \left({\lvert \mathbf{H}^{\mathrm{T}} \rvert}^{\circ2} \bm{\nu}_{\mathbf{s}}^{(t)} \right), \label{eq:vec_amp3} \\
    \mathbf{r}^{(t)} &= \hat{\mathbf{x}}^{(t-1)} + \bm{\nu}_{\mathbf{r}}^{(t)} \odot \left(\mathbf{H}^{\mathrm{T}}\mathbf{s}^{(t)}\right), \label{eq:vec_amp4}
\end{align}
\end{subequations}
where $\mathbf{1} \in \mathbb{R}^{2MN\times 1}$, and the element-wise multiplication and division are represented by operators ``$\oslash$'' and ``$\odot$'', respectively. Besides, we use ${\lvert \mathbf{R} \rvert}^{\circ 2}$ to denote the element-wise magnitude square operation for matrix $\mathbf{R}$. We observe that one matrix-vector multiplication needs to be performed in \eqref{eq:vec_amp0}, \eqref{eq:vec_amp2}, \eqref{eq:vec_amp3}, and \eqref{eq:vec_amp4}. Also, there are $9$ vector operations for \eqref{eq:vec_amp0}-\eqref{eq:vec_amp4}, each of which requires $2MN$ FLOPs. Thus, each iteration of the AMP module requires $4Q+2MN\times 9$ FLOPs, and the AMP module incurs
    \begin{align}
        \text{FLOPs}^{\text{AMP}}_{P=4} &= (4Q+2MN\times9) T \nonumber \\
        &= (4\times 180,224 + 18 \times 1024)\times 15
        \nonumber \\& = 11,089,920
        \end{align}
and
    \begin{align}
        \text{FLOPs}^{\text{AMP}}_{P=8} &= (4Q+2MN\times9) T \nonumber \\
        &= (4\times 360,448 + 18 \times 1024)\times 15 \nonumber \\& = 21,903,360
    \end{align}
FLOPs over $T=15$ iterations for $P=4$ and $8$, respectively.

\subsection{FLOP Counting for the GNN Module}

\subsubsection{Propagation Step}
\label{subsubsec:propagate}
In the \textbf{Propagation Step}, the GNN module performs message aggregation in parallel at each node, i.e., $\sum\nolimits_{j\in {\tilde{\mathcal{N}}}^{(t)}(i)}\mathbf{m}_{j \rightarrow i}^{(t, \ell)}$, which yields $2MN$ aggregated messages in $(2N_u+1)$-dimension. Denote the number of node-pairs in the pair-wise MRF as $N_s^{(t)} \triangleq \sum\nolimits_{i}\lvert {\tilde{\mathcal{N}}^{(t)}}(i) \rvert$. The number of FLOPs required for the addition operations is given as follows:
\begin{align}
&\sum_{i}\sum_{j\in {\tilde{\mathcal{N}}^{(t)}}(i)}\left(N_u+1\right) -2MN(N_u+1) \nonumber \\  
=& N_s^{(t)}(N_u+1) - 2MN(N_u+1) \nonumber \\
=& \left(N_s^{(t)}-2MN \right)  (N_u+1).
\end{align}
Thus, the number of FLOPs required for message aggregation in the proposed AMP-GNN-based detector is given as follows:
\begin{align}
	\text{FLOPs}_{\text{AMP-GNN}}^{\text{AGG}}  &= \sum_{t=1}^{T} \left(N_s^{(t)}-2MN\right) (N_u+1)  L \nonumber \\
	& = \left(\sum_{t=1}^{T} N_s^{(t)}-2MNT \right) (N_u+1) L.
	\label{FLOP:AGG}
\end{align}
Note that the exact value of $\text{FLOPs}_{\text{AMP-GNN}}^{\text{AGG}}$ depends on $N_{s}^{(t)}$, which is related to the channel realization. Hence, the complexity analysis in Section III-D is based on the worst-case
scenario, where $N^{(t)}_s = 2MN[2P(P-1) + 1] \approx 4MNP^2$ and $N^{(t)}_s = 2MN[2P(P-1)(4N_o+1)+8N_o+1] \approx 8MN(2N_o+1)P^2$ with and without IDI approximation, respectively. Detailed analysis of $N_{s}^{(t)}$ in the worst-case scenario is provided in Appendix \ref{appendix:wc}.
Since $\text{FLOPs}_{\text{AMP-GNN}}^{\text{AGG}}$ is linear with respect to $N_{g}\triangleq \sum_{t=1}^{T} N^{(t)}_{s}$, in order to obtain its mean value over a set of channel realizations, it suffices to apply the average value of $N_{g}$, denoted as $\overline{N}_{g}$, for evaluating \eqref{FLOP:AGG}. 

With the proposed IDI approximation scheme, we have $\overline{N}_g =753,525$ for $P=4$ and $\overline{N}_g={2,837,310}$ for $P=8$, which are computed according to the respective 50,000 trials. Hence, the numbers of FLOPs required for message aggregation in the proposed AMP-GNN-based detector with $P=4$ and $8$ are given respectively as
\begin{align}
	\text{FLOPs}_{P=4, \text{AMP-GNN}}^{\text{AGG}}  &= (\overline{N}_g-2MNT) (N_u+1) L \nonumber\\ 
	&= {722,805} \times 9 \times 2  \nonumber \\ 
	&= {13,010,490}.
\end{align}
and
\begin{align}
	\text{FLOPs}_{P=8, \text{AMP-GNN}}^{\text{AGG}} &= (\overline{N}_g-2MNT)  (N_u+1)  L \nonumber\\ 
	& = {2,806,590} \times 9 \times 2  \nonumber\\ 
	&= {50,518,620}.
\end{align}

Without IDI approximation, we have $\overline{N}_g={10,713,900}$ ($P=4$) and $\overline{N}_g={19,612,620}$ ($P=8$). Thus, the required numbers of FLOPs for message aggregation in AMP-GNN-v2 with $P=4$ and $8$ are given respectively as
\begin{align}
    \text{FLOPs}_{P=4, \text{AMP-GNN-v2}}^{\text{AGG}}  &= (\overline{N}_g-2MNT)  (N_u+1)  L \nonumber\\ 
    &= {10,683,180} \times 9 \times 2 \nonumber \\ 
    &= {192,297,240}
\end{align}
and
\begin{align}
    \text{FLOPs}_{P=8, \text{AMP-GNN-v2}}^{\text{AGG}} &= (\overline{N}_g-2MNT)  (N_u+1)  L \nonumber\\ 
    & = {19,581,900} \times 9 \times 2 \nonumber\\ 
    &= {352,474,200}.
\end{align}

The $2MN$ aggregated $(2N_u+1)$-dimensional message vectors are passed to MLP $f_{\bm{\theta}}(\cdot)$ for encoding. If $f_{\bm{\theta}}(\cdot)$ is a three-layer MLP with two hidden-layer dimensions as $N_{h_1}$ and $N_{h_2}$, and output dimension $N_u$, the computational complexity of message encoding is given as follows:
\begin{align}
    \text{FLOPs}^{\text{MLP}}&=2MN\left[(2N_u+1)N_{h_1}+N_{h_1}N_{h_2}+N_{h_2}N_u\right] LT \nonumber\\
    &=2048\times(17\times16 + 16\times 12+ 12\times 8)\times 30 \nonumber\\
    &=34,406,400.
\end{align}

\subsubsection{Update Step}
It is straightforward that \eqref{eq:update_step_2} requires $\text{FLOPs}^{\text{Linear}} = 2MNN_u N_h L T=5,898,240$ FLOPs. To analyze the complexity of \eqref{eq:update_step_1}, $f_{\bm{\varphi}(\cdot)}$ is expanded as follow:
\begin{subequations}
\label{eq:gru}
    \begin{align}
    \mathbf{R}^{(t, \ell)} &= \sigma\left( \mathbf{W}_{xr} \mathbf{X}_{in,i} + \mathbf{W}_{hr} \mathbf{s}^{(t, \ell-1)}_i + \mathbf{b}_r\right), \\
    \mathbf{Z}^{(t, \ell)} &= \sigma\left( \mathbf{W}_{xz} \mathbf{X}_{in,i} + \mathbf{W}_{hz} \mathbf{s}^{(t, \ell-1)}_i + \mathbf{b}_z\right), \\
    \Tilde{\mathbf{s}}^{(t, \ell)}_i &= \text{tanh}\left( \mathbf{W}_{xh} \mathbf{X}_{in,i} + \mathbf{W}_{hh} (\mathbf{R}^{(t, \ell)} \odot \mathbf{s}^{(t, \ell-1)}_i) + \mathbf{b}_h\right), \\
    {\mathbf{s}}^{(t, \ell)}_i &= \mathbf{Z}^{(t, \ell)} \odot \mathbf{s}^{(t, \ell-1)}_i + (1-\mathbf{Z}^{(t, \ell)}) \odot \Tilde{\mathbf{s}}^{(t, \ell)}_i,
\end{align}
\end{subequations}
where $\mathbf{X}_{in,i}=\left[\mathbf{m}_i^{(t, \ell)}, \mathbf{a}_i^{(t)}\right] \in \mathbb{R}^{(N_u+2) \times 1}$, $\mathbf{W}_{xr}$, $\mathbf{W}_{xz}$, $\mathbf{W}_{xh} \in \mathbb{R}^{N_h\times(N_u+2)}$, and $\mathbf{W}_{hr}$, $\mathbf{W}_{hz}$, $\mathbf{W}_{hh} \in \mathbb{R}^{N_h\times N_h}$. The computational complexity of \eqref{eq:gru} is given as $3[N_h  (N_u+2)+ N_h^2] + 11N_h$, and therefore the number of FLOPs required for the GRU can be expressed as follows:
\begin{align}
    \text{FLOPs}^{\text{GRU}} &= 2MN \times \left\{ 3[N_h (N_u+2)+ N_h^2] + 11N_h \right\} LT \nonumber\\
    &= 2048 \times \left[ 3(12\times 10 + 12\times 12)+11\times 12 \right] \times 30 \nonumber\\
    & = 24,330,240.
\end{align}
Hence, the total number of FLOPs required for the \textbf{Update Step} is given as $\text{FLOPs}^{\text{Update}}=\text{FLOPs}^{\text{Linear}}+\text{FLOPs}^{\text{GRU}}=30,228,480$.

\subsubsection{Readout Step}
The computational complexity of the \textbf{Readout Step} is determined by MLP $f_{\bm{\omega}}(\cdot)$. If $f_{\bm{\omega}}(\cdot)$ is a three-layer MLP with hidden-layer dimensions $N_{h_1}$ and $N_{h_2}$, and output dimension $|\mathcal{Q}_{\text{R}}|$, and the number of FLOPs required can be calculated as follows:
\begin{align}
    \text{FLOPs}^{\text{Readout}}&=2MN(N_uN_{h_1}+N_{h_1}N_{h_2}+N_{h_2}|\mathcal{Q}_{\text{R}}|)T \nonumber\\
    &=2048\times(8\times16 + 16\times 12+ 12\times 4)\times 15 \nonumber\\
    &= 11,304,960.
\end{align}

\subsection{Overall Computational Complexity}
Given the detailed analysis for each processing step of the AMP-GNN-based detector, we summarize the overall computational complexity as follows:
\begin{align}
    \text{FLOPs}_{\text{AMP-GNN}} &= \text{FLOPs}^{\text{AMP}} + \text{FLOPs}^{\text{AGG}}+ \text{FLOPs}^{\text{MLP}}  \nonumber \\
    &  + \text{FLOPs}^{\text{Update}} + \text{FLOPs}^{\text{Readout}}.
    \label{eq:FLOP_Total}
\end{align}

For $P=4$, we substitute $\text{FLOPs}^{\text{AMP}}$ with $\text{FLOPs}^{\text{AMP}}_{P=4}$ and $\text{FLOPs}^{\text{AGG}}$ with $\text{FLOPs}_{P=4, \text{AMP-GNN}}^{\text{AGG}}$ in \eqref{eq:FLOP_Total}, the number of FLOPs required by the AMP-GNN-detector is given as
\begin{align}
	&{\text{FLOPs}_{P=4, \text{AMP-GNN}}} = 11,089,920 + {13,010,490} \nonumber \\
	&+ 34,406,400 +  30,228,480 + 11,304,960 \nonumber \\
	& = {100,040,250}.
\end{align}
For $P=8$, we substitute $\text{FLOPs}^{\text{AMP}}$ with $\text{FLOPs}^{\text{AMP}}_{P=8}$ and $\text{FLOPs}^{\text{AGG}}$ with $\text{FLOPs}_{P=8, \text{AMP-GNN}}^{\text{AGG}}$ in \eqref{eq:FLOP_Total}, the number of FLOPs required by the AMP-GNN-detector is given as
\begin{align}
	& {\text{FLOPs}_{P=8, \text{AMP-GNN}}} = 21,903,360 + {50,518,620}  \nonumber \\
	&+ 34,406,400 + 30,228,480 + 11,304,960 \nonumber \\
	& = {148,361,820}.
\end{align}

Similarly, we obtain the numbers of FLOPs required by the AMP-GNN-v2 with $P=4$ and $P=8$ respectively as
\begin{align}
    & {\text{FLOPs}_{P=4, \text{AMP-GNN-v2}}} = 11,089,920 + {192,297,240}  \nonumber \\
    & + 34,406,400 + 30,228,480 + 11,304,960 \nonumber \\
    & = {279,327,000}
\end{align}
and
\begin{align}
    & {\text{FLOPs}_{P=8, \text{AMP-GNN-v2}}} = 21,903,360 + {352,474,200} \nonumber \\
    &+ 34,406,400 + 30,228,480 + 11,304,960 \nonumber \\
    & = {450,317,400}.
\end{align}

\section{Number of node-pairs in the worst-case scenario}
\label{appendix:wc}
\begin{figure}[h]
	\centering
	\includegraphics[width=\linewidth]{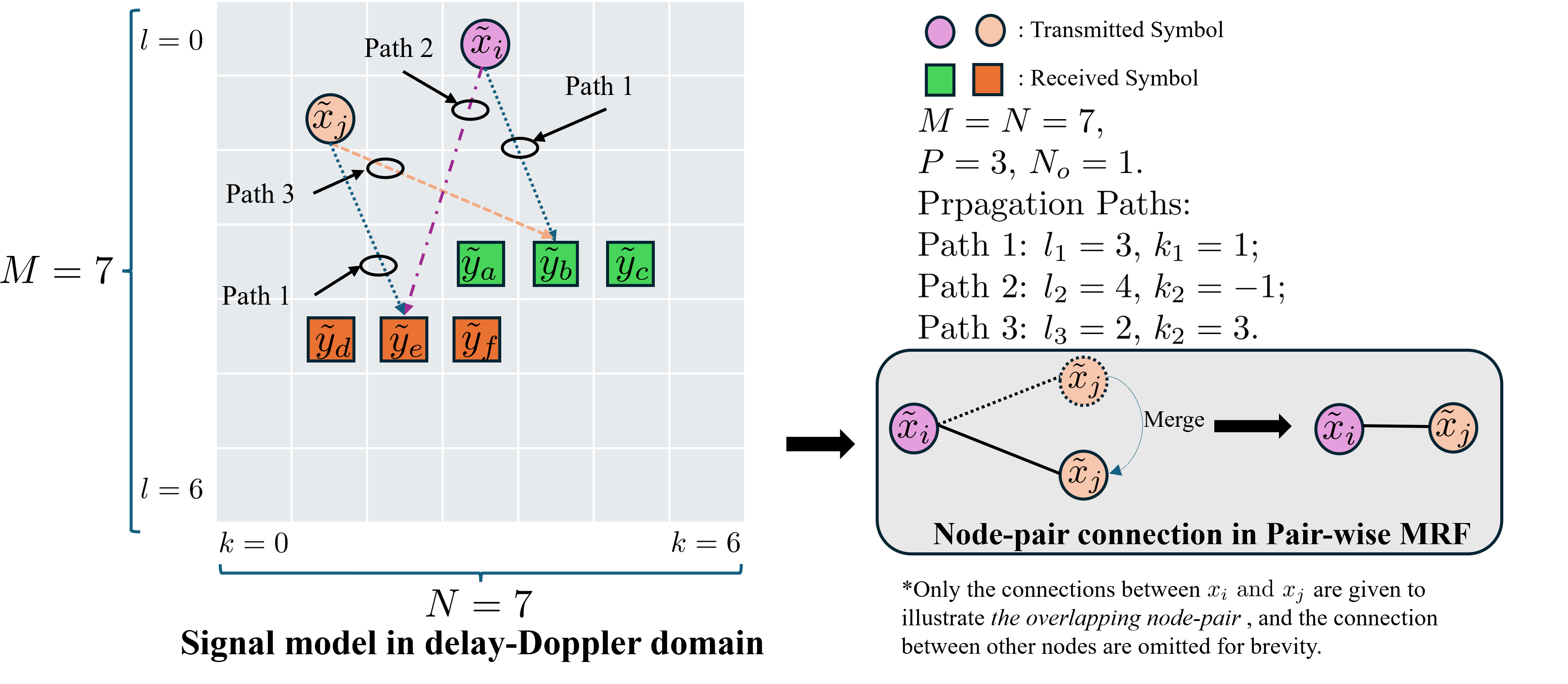}
	\caption{.~~An example of overlapping node-pairs, where $l_p$ and $k_p, p=1, 2, 3$ denote the integer delay and Doppler shift indexes, respectively.}
	\label{fig:overlapping}
\end{figure}

\begin{figure}[h]
	\centering
	\includegraphics[width=\linewidth]{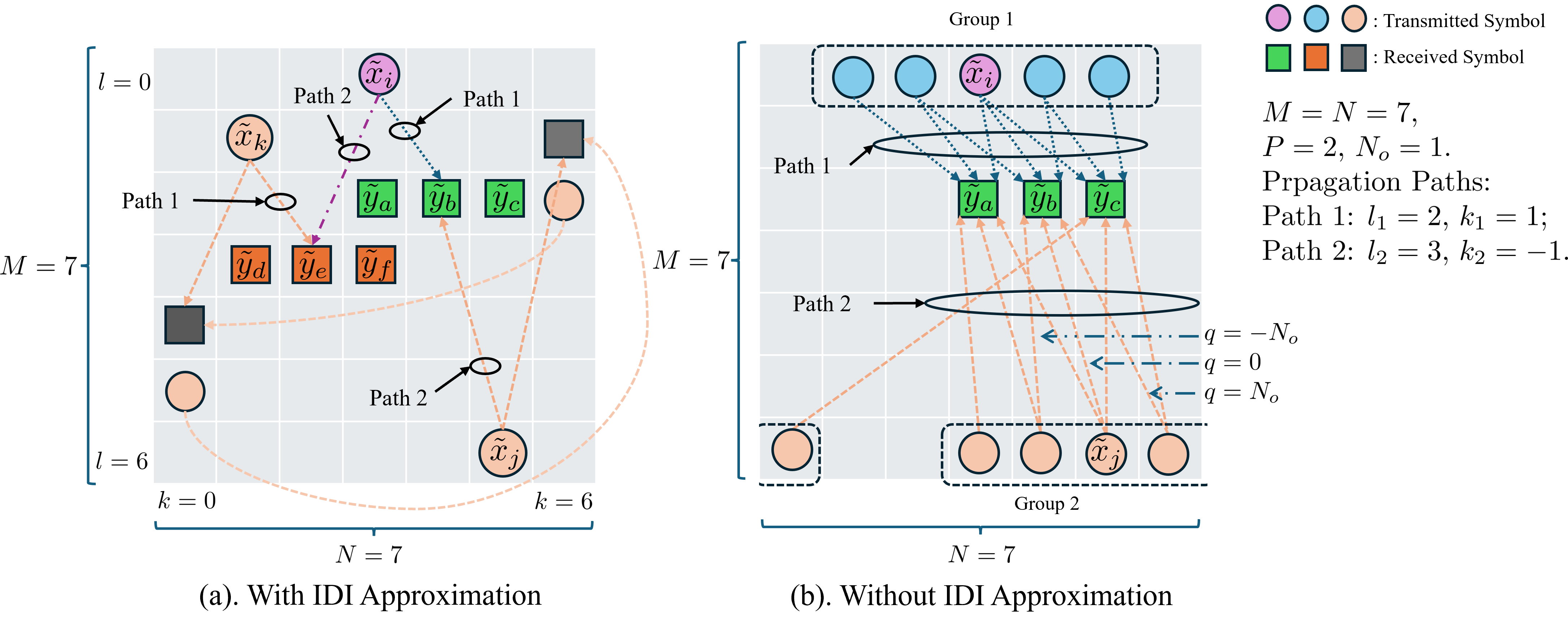}
	\caption{.~~Illustrative examples of the signal model with and without IDI approximation in the delay-Doppler domain, where $l_p$ and $k_p, p=1, 2$ denote the integer delay and Doppler shift indexes, respectively.}
	\label{fig:toy_model}
	\vspace{-1em}
\end{figure}

In the worst-case scenario, the node-pairs associated with different propagation paths in the pair-wise MRF are assumed to be non-overlapping. To illustrate this, we consider a signal model $\tilde{\mathbf{y}}=\tilde{\mathbf{H}} \tilde{\mathbf{x}}$ with $\tilde{x}_m$ and $\tilde{y}_m$ respectively denote the $m$-th entry of $\tilde{\mathbf{x}}$ and $\tilde{\mathbf{y}}$. An example of overlapping node-pairs is shown in Fig. \ref{fig:overlapping}, where the two transmitted symbols $\tilde{x}_i$ and $\tilde{x}_j$ are connected by the received symbol $\tilde{y}_b$ through Path $1$ and Path $3$, respectively. Meanwhile, they are also connected by the received symbol $\tilde{y}_e$ through Path $2$ and Path $1$, respectively. Denote $\tilde{\mathbf{h}}_m$ as the $m$-th column of $\tilde{\mathbf{H}}$. In this case, there are more than one entries, i.e., the $b$-th and the $e$-th entry, of both $\tilde{\mathbf{h}}_i$ and $\tilde{\mathbf{h}}_j$ being non-zero, and thus node $\tilde{x}_i$ and $\tilde{x}_j$ are connected in the pair-wise MRF. The node-pairs between $\tilde{x}_i$ and $\tilde{x}_j$ overlap as there are more than one entries of $\tilde{\mathbf{h}}_i$ and $\tilde{\mathbf{h}}_j$ at the same position being non-zero, as shown at the bottom right of Fig. \ref{fig:overlapping}. It can be verified that the non-overlapping cases can achieve the largest number of node-pairs in the pair-wise MRF. To derive the number of node-pairs in the worst-case scenario, illustrative examples with and without IDI approximation are provided in Fig. \ref{fig:toy_model}. 

With IDI approximation, it can be observed from the signal model in Fig. \ref{fig:toy_model}(a) that each transmitted symbol is only associated with one received symbol through each propagation path, and the received symbol is associated with other $P-1$ transmitted symbols. Thus, each transmitted symbol is connected with other $P-1$ transmitted symbols in the pair-wise MRF through each propagation path. For example, $\tilde{x}_i$ is associated with $\tilde{y}_b$ through Path 1, and $\tilde{y}_b$ is associated with $\tilde{x}_j$ through another propagation path, i.e., Path 2. Since there are $P$ propagation paths, each node (e.g., $\tilde{x}_i$), has a total number of $P(P-1)$ (i.e., $2\times 1 = 2$) neighboring nodes (i.e., $\tilde{x}_j, \tilde{x}_k$) in the pair-wise MRF shown in Fig. \ref{fig:toy_model}(a). Thus, we obtain $N_s = 2MN[2P(P-1)+1]$ in the real-valued pair-wise MRF with IDI approximation. 

Without IDI approximation, it can be observed from the signal model in Fig. \ref{fig:toy_model}(b) that each transmitted symbol (e.g., $\tilde{x}_i$) is associated with $2N_o+1$ received symbols (i.e., $\tilde{y}_a$, $\tilde{y}_b$, $\tilde{y}_c$) through each propagation path. These received symbols are further associated with $P-1$ groups of symbols (e.g., Group 2 as shown in Fig. \ref{fig:toy_model}(b)), with a total number of $4N_o+1$ symbols, through other propagation paths. Thus, in the corresponding pair-wise MRF, for $P$ propagation paths, the total number of neighboring nodes belonging to different group per node is $P(P-1)(4N_o+1)$, while that within the same group (i.e., Group 1) is $4N_o$. Hence, each node has a total number of $P(P-1)(4N_o+1)+4N_o$ neighboring nodes in the corresponding pair-wise MRF as shown in Fig. \ref{fig:toy_model}(b). Thus, we obtain $N_s=2MN[2P(P-1)(4N_o+1)+8N_o+1]$ in the real-valued pair-wise MRF without IDI approximation.

\section{FLOPs and Runtime Decomposition}
\label{sec:FLOP_Count_Table}
Table \ref{tab:flops} and \ref{tab:time} respectively summarize the required number of FLOPs and runtime of three OTFS data detectors, including the GNN-based detection, AMP-GNN-based detector, and AMP-GNN-v2. These values are decomposed to the key processing steps.

\begin{table}[h]
	\centering
	\begin{tabular}{l|c|c|c|c|c|c}
		\hline
		\multirow{2}{*}{Processing} & \multicolumn{2}{c|}{GNN [7]} & \multicolumn{2}{c|}{AMP-GNN-v2} & \multicolumn{2}{c}{AMP-GNN}\\
		\cline{2-7}
		& $P=4$& $P=8$ & $P=4$ & $P=8$ & $P=4$ & $P=8$ \\
		\hline \hline
		Aggregation & $8.57$  & $15.69$ & $19.23$ & $35.25$ & $1.30$  & $5.05$  \\
		MLP $f_{\bm{\theta}}(\cdot)$       & $8.26$ & $8.26$ & $3.44$ & $3.44$ & $3.44$ & $3.44$ \\
		Update     & $1.53$ & $1.53$ & $3.02$ & $3.02$ & $3.02$ & $3.02$ \\
		Readout     & $0.08$ & $0.08$ & $1.13$ & $1.13$ & $1.13$ & $1.13$  \\
		AMP         & $0$ & $0$ & $1.11$ & $2.19$ & $1.11$ & $2.19$  \\\hline
		Total & $18.44$ & $25.56$ & $27.93$ & $45.03$ & $10.00$ & $14.84$ \\ \hline
	\end{tabular}
	\caption{.~~FLOPs ($\times10^{7}$) decomposition.}
	\label{tab:flops}
\end{table}

\begin{table}[h]
	\centering
	\begin{tabular}{l|c|c|c|c|c|c}
		\hline
		\multirow{2}{*}{Processing} & \multicolumn{2}{c|}{GNN [7]} & \multicolumn{2}{c|}{AMP-GNN-v2} & \multicolumn{2}{c}{AMP-GNN}\\
		\cline{2-7}
		& $P=4$& $P=8$ & $P=4$ & $P=8$ & $P=4$ & $P=8$ \\
		\hline \hline
		Initialization & $9.4$ & $17.0$ & $8.8$ & $14.9$ & $6.7$ & $6.8$ \\
		Aggregation & $7.5$  & $9.5$ & $12.1$ & $14.3$ & $4.5$  & $4.9$  \\
		MLP  $f_{\bm{\theta}}(\cdot)$      & $5.9$ & $6.6$ & $5.5$ & $5.6$ & $4.8$ & $5.0$ \\
		Update      & $2.0$ & $2.1$ & $3.7$ & $3.7$ & $3.5$ & $3.5$ \\
		Readout     & $1.3$ & $1.3$ & $2.5$ & $2.5$ & $2.2$ & $2.3$  \\
		AMP        & $0$ & $0$ & $5.1$ & $5.6$ & $5.1$ & $5.6$  \\\hline
		Total & $26.1$ & $36.5$ & $37.7$ & $46.6$ & $26.8$ & $28.1$ \\ \hline
	\end{tabular}
	\caption{.~~Runtime (ms) decomposition.}
	\label{tab:time}
\end{table}

 It can be observed that for both the GNN-based detector and AMP-GNN-v2, message aggregation contributes to a significant proportion of the computational overhead compared to other processing steps. However, this is not the case for the AMP-GNN-based detector, where all the processing steps have comparable computational complexity. Specifically, compared to AMP-GNN-v2, the number of FLOPs performed in the message aggregation step of the AMP-GNN-based detector is reduced by 93.2\% for $P=4$ and 85.7\% for $P=8$. Likewise, 62.8\% and 65.7\% of the runtime corresponding to the message aggregation step is saved for $P=4$ and $8$, respectively. These results further justify the effectiveness of the proposed IDI approximation scheme in reducing the computational overhead.

\end{document}